\documentclass[aps,twocolumn,amsmath,amsfonts,amssymb]{revtex4}
\usepackage{graphicx}
\usepackage{epsf}

\begin{document}
\title{Mesoscopic Quantum Superposition of Generalized Cat State: A Diffraction limit}

\author{Suranjana Ghosh$^{\mathrm{1}}$\footnote{e-mail:sghosh@iitp.ac.in},
Raman
Sharma$^{\mathrm{2}}$\footnote{e-mail:ramansharma@cse.iitb.ac.in},
and Prasanta K.
Panigrahi$^{\mathrm{1}}$\footnote{e-mail:pprasanta@iiserkol.ac.in}}

\affiliation{$^{\mathrm{1}}$ Indian Institute of Science Education
and Research Kolkata, Mohanpur - 741246, India
\\ $^{\mathrm{2}}$Indian Institute of Technology Bombay, Powai, Mumbai - 400076,
India}

\begin{abstract}
The orthogonality of cat and displaced cat states, underlying
Heisenberg limited measurement in quantum metrology, is studied in
the limit of large number of states. The mesoscopic superposition
of the generalized cat state is correlated with the corresponding
state overlap function, controlled by the sub-Planck structures
arising from phase space interference. The asymptotic expression
of this overlap function is evaluated and the validity of large
phase space support and distinguishability of the constituent
states, in which context the asymptotic limit is achieved, are
discussed in detail. For large number of coherent states,
uniformly located on a circle, the overlap function significantly
matches with the diffraction pattern for a circular ring source
with uniform angular strength. This is in accordance with the van
Cittert-Zernike theorem, where the overlap function, similar to
the mutual coherence function, matches with a diffraction pattern.
The physical situation under consideration is delineated in phase
space by utilizing Husimi-Q function.
\end{abstract}

\maketitle

\section{Introduction}

Cat states and their generalizations are known to achieve
Heisenberg limited sensitivity in estimation of parameters like
coordinate/momentum displacements and phase space rotations
\cite{Zurek}. A criterion to distinguish quantum states without
classical counterparts, from those not possessing the same, are
studied in \cite{Vogel,Richter}. For these non-classical states,
subtle interference effects in the phase space \cite{Schleich
Wheeler} lead to sub-Planck structures in their Wigner functions,
which in turn allow precision measurement of quantum parameters,
bettering the standard quantum limit. Recently, sub-Planck
structures in different physical systems have been investigated
\cite{GSAg,Jay,Toscano,bhatt,ghosh,Roy,sghosh1,sghosh2}. It has
been demonstrated \cite{Toscano,ghosh,Roy} that, the sensitivity
of the state used in quantum metrology is directly related to the
area of the sub-Planck structures: $ \rho =\frac{\hbar^2}{A}$,
with $A$ being the action of the effective support of the Wigner
function. The interference in phase space is a pure quantum
phenomenon, arising due to the fact that these states are
superposition of the coherent states (CSs), which themselves are
classical. The increase in the number of interfering coherent
states in the phase space is akin to emergence of diffraction in
classical optics, when the number of interfering sources becomes
large with sufficient phase space support.

Here, we analyze this diffraction limit of the smallest
interference structures and find an exact asymptotic value of the
displacement sensitivity. With the assumption of large phase-space
support for the estimating state and smallness of the quantum
parameters to be estimated, it is found that the asymptotic limit
of the sensitivity reaches $|\delta| = \frac{C}{2 |\alpha|}$,
where $C$ is the first root of $J_0$, the $0^{th}$ order Bessel
function. We explicitly show that this assumption is adequate for
realistic values of the physical parameters; \emph{i.e.}, the
average photon number and the number of superposed CSs. The
numerical analysis depicts how the asymptotic limit of exact
overlap function (OF) reaches to the $0^{th}$ order Bessel
function for higher order mesoscopic superpositions. This limiting
behavior in the phase space interference is found to be analogous
to the van Cittert-Zernike theorem \cite{Wolf}, relating the
mutual coherence in classical optics to diffraction. A phase space
distribution (Q-function), having only positive regions, reveals
the actual physical situation at the point of resemblance between
the two theories.

\section{Results and Discussions}

Cat states and their generalizations play a significant role in
quantum optics and quantum computation \cite{Milburn}. A number of
experimental schemes exist to produce cat states in laboratory
conditions \cite{Haroche}. These ``pointer states''\cite{pointer}
often naturally manifest, when suitable quantum systems are
coupled with decohering environment. It has been observed that the
robustness of these states, made out of classical CSs, is a result
of ``quantum Darwinism'' \cite{Darwin}. We consider a single
oscillator, with the CS being an eigen state of $a$:
$a|\alpha\rangle = \alpha |\alpha\rangle$, with annihilation and
creation operator $a$ and $a^{\dagger}$ : $[a,a^{\dagger}]=1$.

The generalized cat state is composed of CSs, equally phase
displaced on a circle:
\begin{equation}\label{cat}
 |cat_{n,\alpha} \rangle = \frac{1}{\sqrt{n}} \sum_{j=1}^{n} |e^{\frac{\iota 2 \pi j}{n} } \alpha \rangle = \frac{1}{\sqrt{n}} \sum_{j=1}^{n} D(e^{\frac{\iota 2 \pi j}{n} } \alpha )|0\rangle \text{,}
\end{equation}
where, $|\alpha\rangle = D(\alpha) |0\rangle$, with the
displacement operator, $D(\alpha)=e^{\alpha a^{\dagger} -
\alpha^{\star} a}$ and $a|0\rangle=0$. Here, it is worth
mentioning that the CSs are assumed to be distinguishable
\cite{normalization}. The displacements in the coordinate and
momenta can be realized through an appropriately displaced cat
state \cite{Toscano}:
$|cat_{n,\alpha}^{\delta}\rangle=D(\delta)|cat_{n,\alpha}\rangle$.

For checking the sensitivity of the estimating state
$|cat_{n,\alpha} \rangle$, one computes the overlap of the same
with the displaced state and studies the orthogonality conditions,
\begin{widetext}
\begin{eqnarray}
\langle cat_{n,\alpha}|cat_{n,\alpha}^{\delta}\rangle
&=& \frac{1}{n} \sum_{j=1}^n \sum_{k=1}^n \langle 0|D(e^{\frac{i 2 \pi j}{n} }
\alpha )^{\dagger} D(\delta) D(e^{\frac{i 2 \pi k}{n} } \alpha ) |0\rangle \nonumber \\
&=& \frac{1}{n} \sum_{j=1}^n \sum_{k=1}^n (e^{i Im(\delta
\alpha^{\star} ( e^{- \frac{i 2 \pi j}{n}}+e^{- \frac{i 2 \pi
k}{n}}) + |\alpha|^2 e^{- \frac{i 2 \pi (k-j)}{n}})}) (e^{-
\frac{1}{2} |\delta + \alpha (e^{- \frac{i 2 \pi k}{n}} - e^{-
\frac{i 2 \pi j}{n}})|^2})
\text{.}\label{over1}\\
&=&\frac{1}{n} \sum_{j=1}^n \sum_{k=1}^n e^{i(2r
cos(\frac{\pi(j-k)}{n})\;sin(\theta-\frac{\pi(j+k)}{n})+|\alpha|^2\;sin(\frac{2\pi(j-k)}{n}))}\nonumber\\
&&\;\;\;\;\;\;\;\;\;\;\times
e^{-\frac{1}{2}(|\delta|^2+2|\alpha|^2(1-cos(\frac{2\pi(j-k)}{n}))+4r\;sin(\frac{\pi(j-k)}{n})sin(\theta-\frac{\pi(j+k)}{n}))}\label{over2}
\end{eqnarray}
\end{widetext}
where, $r = |\alpha||\delta|$ and $\theta = (\theta_\delta
-\theta_\alpha)$ with $\alpha = |\alpha|e^{i \theta_\alpha}$ and
$\delta = |\delta|e^{i \theta_\delta}$.

\begin{figure}[htbp]
\centering
\includegraphics[width=3in]{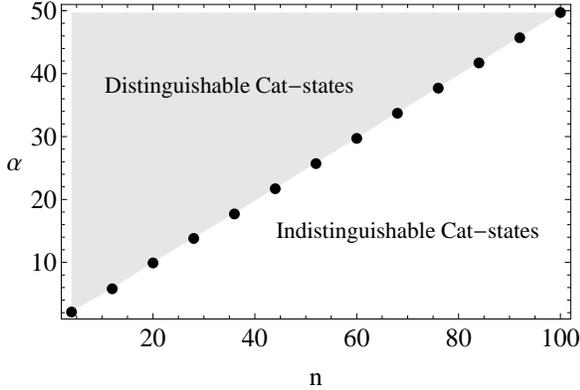}
\caption{Parameter domain of the phase space support $\alpha$ to
maintain the condition of distinguishability. The shaded region
designates the domain for distinguishable
cat-states.}\label{alphan}
\end{figure}

The phase space of the generalized cat state of Eq.~(\ref{cat}) is
composed of `$n$' CSs, equally placed in a circle of radius
$|\alpha|$, where large phase space support means the large
magnitude of $|\alpha|$. Now the natural question arises, how
large should be the phase space support to maintain the
distinguishability for a given mesoscopic superposition, {\it
i.e.}, domain of $\alpha$ for a given $n$. This is delineated in
Fig.~\ref{alphan}, where we have taken upto a large value of $n$,
({\it e.g.} $n=100$). The upper region of the plot (shaded)
depicts the allowed parameter domain of $\alpha$ for mesoscopic
superposition of CSs. Table-I reveals the values of alpha
(accurate upto first decimal place), above which the states are
distinct. We further observe that the maximum value of the ratio
$n/\alpha$ to maintain distinguishability takes the average value
$\nu = 2.016$. Thus one does not need a very large $\alpha$ to
generate the said interference structures in phase space. In fact,
increasing $n$ is quite difficult in experiments, as it requires a
large nonlinearity of the medium. On the contrary, the absolute
value of $\alpha$ is directly related to the average photon number
of the coherent state, which can be manipulated by controlling the
laser beam. Hence, the allowed maximum order of mesoscopic
superposition (`$n$') for a given $\alpha$, conforming our result,
is sufficiently large in reality.

\begin{table*}\centering
\caption{Estimate of the numerically obtained minimum value of
phase space area (proportional to $|\alpha|$), required for a
mesoscopic superposition of $n$- CSs.} \vskip .1in
\begin{tabular} {|c||c|c|c|c|c|c|c|c|c|c|c|c|c|}
\hline $\;\;\;\;\;n\;\;\;\;\;$ &  4  &  12  &  20  &  28  &  36  &  44  &  52  &  60  &  68  &  76  &  84  &  92  &  100  \\
\hline $\alpha$ & 2.1 & 5.8 & 9.9 & 13.8 & 17.7 & 21.7 & 25.7 & 29.7 & 33.7 & 37.7 & 41.7 & 45.7 & 49.7 \\
\hline
\end{tabular}
\label{table} \vskip .01in
\end{table*}

It is important to mention that, the OF between the initial and
displaced cat states can also be represented by the phase space
Wigner distribution:
\begin{equation}
|\langle
cat_{n,\alpha}|cat_{n,\alpha}^{\delta}\rangle|^2=\int\!\!\int
W_{cat_{n,\alpha}}(x,p) W_{cat_{n,\alpha}^{\delta}}(x,p)
dx\;dp.\label{wigoverlap}
\end{equation}
This relation reveals the physical significance of the
oscillations of the OF in a particular direction in phase space
and connects our result with the mesoscopic superposition
structures. The oscillation of the OF is the signature of quantum
interference structures of dimension less that Planck's constant,
{\it i.e.}, sub-Planck scale structures. Each zero of the OF
signifies the orthogonality of the original and displaced states,
thereby implying the sensitivity limit of Heisenberg limited
measurement.

Now, it is intuitive as well as numerically verified by us that
the entire contribution of the OF in Eq.~(\ref{over1}) or
(\ref{over2}) mainly originates from the adjacent components of
the original and displaced cat states, {\it i.e.}, $j\sim k$.
Therefore, $|j-k|<<n$, $cos(\pi(j-k)/n)\rightarrow 1$ and
$sin(\pi(j-k)/n)\rightarrow 0$. Then Eq.~(\ref{over2}) takes the
simpler form
\begin{equation}
\langle cat_{n,\alpha}|cat_{n,\alpha}^{\delta}\rangle =
\frac{e^{-\frac{1}{2}|\delta|^2}}{n} \sum_{j=1}^n \sum_{k=1}^n
cos\left[2r sin(\theta-\frac{\pi(j+k)}{n})\right].\label{over4}
\end{equation}
The off-diagonal terms in the above expression have negligible
contribution. This assumption bears similar meaning of the
classical situation, where an incoherent ring source is assumed,
for which the cross-correlations between the different points of
the source can be neglected. Now, with the assumption of
sufficient phase-space support for the estimating state and
smallness of quantum parameters to be estimated, one can consider
only the diagonal terms and obtain
\begin{eqnarray}
\langle cat_{n,\alpha}|cat_{n,\alpha}^{\delta}\rangle &\approx&
\frac{e^{-\frac{1}{2}|\delta|^2}}{n} \sum_{j=1}^{n} cos\left[2r
  sin(\theta-\frac{2\pi j}{n})\right] \nonumber\\
  &\approx& \frac{1}{n} \sum_{j=1}^{n} cos\left[2r sin(\theta-\frac{2\pi
j}{n})\right]\text{.}
\end{eqnarray}
It needs to be mentioned that the state overlap depends only on
$\delta \alpha^{\star}$, which leads to the conclusion that the
sensitivity of estimating $\delta$ is inversely proportional to
$|\alpha|$. It is easily checked that the OF, being of
interferometric origin, is only sensitive to the difference in
phase : $\langle
cat_{n,\alpha}^{\delta_2}|cat_{n,\alpha}^{\delta_1}\rangle = e^{i
\phi}  \langle cat_{n,\alpha}|cat_{n \alpha}^{\delta_1 -
\delta_2}\rangle$. The OF for $n=2$,
\begin{equation}
|\langle cat_{2,\alpha}|cat_{2,\alpha}^{\delta}\rangle|^2 \approx
\cos^2 (2 |\alpha|\delta_\perp),
\end{equation}
\begin{figure}
\centering
\includegraphics[width=2.5in]{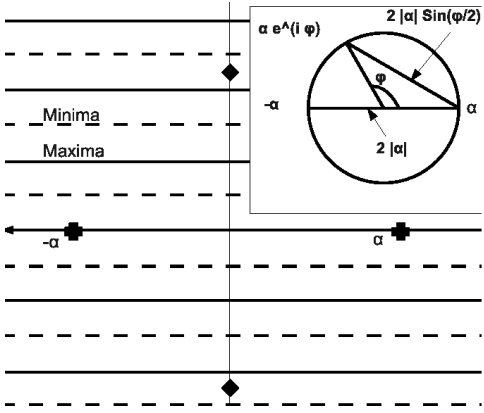}
\caption{Analogy between two-source interference: The solid lines
show the maximum and dashed lines show the minimum intensity
values. The crosses are the positions of coherent states and
diamonds show the equivalent positions of sources of light which
will produce the same pattern at a distance. The inset shows the
equivalent position of sources for the state
$|cat2_{\alpha,\phi}\rangle = \frac{|\alpha\rangle+|\alpha
e^{\iota \phi}\rangle}{2}$} \label{figure}
\end{figure}
which matches with the known result \cite{Toscano}, with \(
\delta_\perp = |\delta| \sin (\theta_\delta -\theta_\alpha) \text{
and } \delta_\parallel = |\delta| \cos (\theta_\delta
-\theta_\alpha)\). As is depicted in Fig.~\ref{figure}, it is
interesting to observe that the above expression is analogous to
the double slit interference pattern, where the normalized
intensity can be written as $\frac{I}{I_{max}} = \cos^2 (\frac{y b
\pi}{s \lambda})$ \cite{Hecht}. The path difference between the
two waves reaching at the observation point is $yb/s$, where $b$
defines the distance between the two slits, $s$ is the separation
between the aperture and the screen, and $y$ corresponds to the
vertical coordinate of the detector. The above analogy can be
mathematically established by taking $\lambda$ in the unit of $s$
and redefining the commutation relation, $[a,a^{\dagger}] = \pi
\lambda^{-1}$:
\begin{equation}
|\langle cat_{2,\alpha}|cat_{2,\alpha}^{\delta}\rangle|^2 = \cos^2
\left[2 \frac{|\alpha|\delta_\perp \pi}{\lambda}\right],
\end{equation}
where $2|\alpha|$ is the separation of the two coherent state
sources. Use of the phase shifted cat state,
$|cat2_{\alpha,\phi}\rangle = \frac{|\alpha\rangle+|\alpha
e^{\iota \phi}\rangle}{2}$, would yield an interference pattern at
an angle $\frac{\phi}{2}$ and \textit{fringe width},
$2|\alpha|\sin\frac{\phi}{2}$:
\begin{equation}
|\langle cat2_{\alpha,\phi}|cat2_{\alpha,\phi}^{\delta}\rangle|^2
=  \cos^2 (2 |\alpha|\sin\frac{\phi}{2} (\delta_\perp
\sin\frac{\phi}{2} + \delta_\parallel \cos\frac{\phi}{2}))
\end{equation}

Introducing a phase between the constituent CSs of a cat state
with $n=2$ gives the state $|cat2_{\alpha}^{\phi}\rangle =
\frac{|\alpha\rangle+e^{\iota \phi}|-\alpha\rangle}{2}$. The OF
for this state is
\begin{equation}
|\langle
cat2_{\alpha}^{\phi}|cat2_{\alpha}^{\phi,\delta}\rangle|^2 =
\cos^2 (2 |\alpha|\delta_\perp) - \phi) \text{,}
\end{equation}
akin to the phenomenon of ``\textit{fringe shift}'' observed in
classical optics.

We now derive the state overlap and sensitivity in parameter
estimation for very high order of mesoscopic superpositions. For
convenience, we assume $n$ is even:
\begin{equation}
\langle cat_{n,\alpha}|cat_{n,\alpha}^{\delta}\rangle =
\frac{2}{n} \sum_{j=1}^{\frac{n}{2}} \cos\left[2 r \sin (\theta -
\frac{2 \pi j}{n})\right].
\end{equation}
In the asymptotic limit of $n$, one writes
\begin{widetext}
\begin{eqnarray}
\lim_{n \rightarrow \infty,\; n/\alpha\leq\nu} \langle
cat_{n,\alpha}|cat_{n,\alpha}^{\delta}\rangle
&=& \lim_{n \rightarrow \infty,\; n/\alpha\leq\nu} \frac{2}{n} \sum_{j=1}^{\frac{n}{2}} \cos (2 r \sin (\theta - \frac{2 \pi j}{n})) \nonumber   \\
&=& \lim_{n \rightarrow \infty,\; n/\alpha\leq\nu} \frac{1}{n} \sum_{j=1}^{n} \cos (2 r \sin (\theta - \frac{2 \pi j}{n})) \nonumber \\
&=& \int_0^1 \cos(2 r \sin (\theta - 2 \pi x)) dx = \frac{1}{2 \pi} \int_{0}^{2 \pi} \cos(2 r \sin (z)) dz \nonumber \\
&=& J_0(2 |\alpha| |\delta|) \text{.}\label{bessel}
\end{eqnarray}
\end{widetext}

\begin{figure}[htpb]
\includegraphics[width=3.6 in]{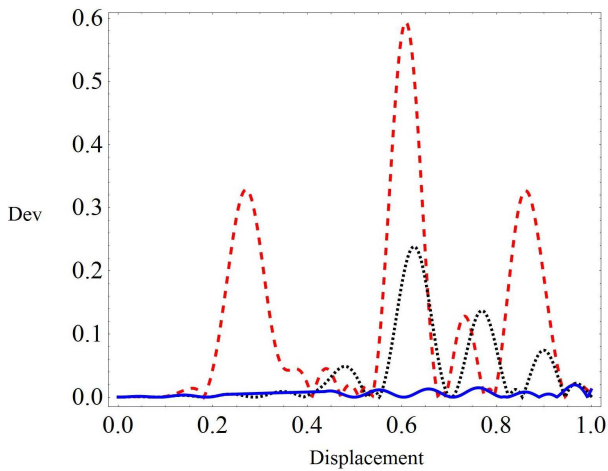}
\centering \caption{Deviation (Dev) of the overlap function
(Eq.~(\ref{over2})) from the zeroth order Bessel function for
$\alpha=15$: $n=8$ (dashed line), $n=16$ (dotted line), and $n=30$
(solid line). For larger value of '$n$', the OF almost coincides
with the zeroth order Bessel function.} \label{overall}
\end{figure}

\begin{figure*}
\centering
\includegraphics[width=6.5in]{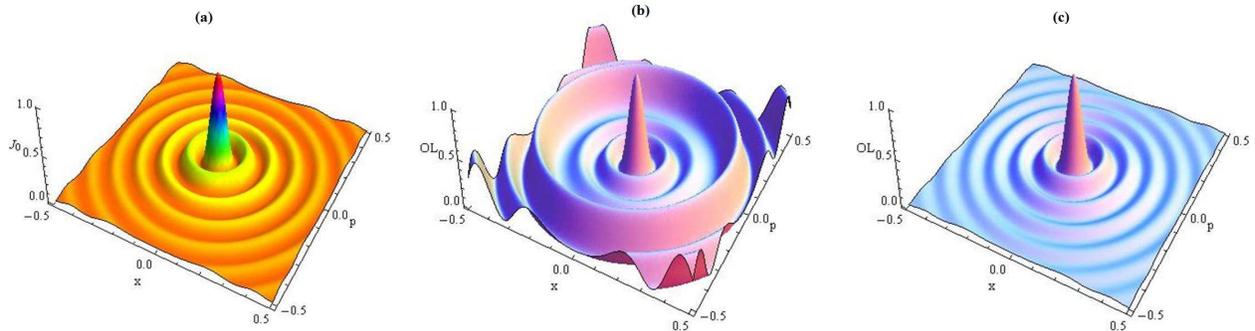}
\caption{a) 3D plot of the Bessel function of zeroth order ($J_0$)
for $\alpha=15$, b) 3D plot of overlap (OL) for $\alpha=15$,
$n=10$, and c) 3D plot of overlap (OL) for $\alpha=15$ for
$n=30$.} \label{plot3D}
\end{figure*}

This proves our assertion that states can be discriminated for
$|\delta| = \frac{C}{2 |\alpha|}$ due to orthogonality, where $C$
is a root of the Bessel function (of first kind) of order zero,
$i.e.,$ $J_0$. In Eq.~\ref{bessel}, we evaluate the overlap
function for very high order of mesoscopic superposition.
Theoretically, a limit $n \rightarrow \infty$ is taken,
provided/implied that the states are still distinguishable. It is
worth mentioning that theoretically the parameter $\alpha$ also
does not have any upper limit and in principle can go upto
infinity. Thus, the Bessel function is obtained without any
further restriction. In reality, neither $n$ nor $\alpha$ can go
upto infinity and thus one has to take the practical quantitative
estimate of these two physical parameters, as described in
Fig.~\ref{alphan} and Table-I. The distinguishability condition
implies $n/\alpha\leq\nu$, where the average value of $\nu$ is
$2.016$.

In Fig.~\ref{overall}, we have plotted the deviation
($Dev=OF-Bessel\; Function$) with respect to the displacement in
phase space, for three different values of $n$ with $\alpha=15$.
The best result is obtained for $n=30$ or for $n/\alpha\approx
\nu$, beyond which the constituent CSs become indistinguishable.
Thus, for higher order mesoscopic superposition, our result fits
very well with the condition when the phase space support is
sufficient enough. The result is verified by a three dimensional
plotting of the functions in Fig.~\ref{plot3D}. The first plot
corresponds to the bessel function for $\alpha=15$ (a) with
respect to the real and imaginary components of the displacement
parameter. The same is also performed for the overlap function.
The OF for $n=10$ (b) does not match with the Bessel function,
whereas the oscillations find a remarkable similarity with the
Bessel function for $n=30$ (c).

\begin{figure*}
\centering
\includegraphics[width=6.5 in]{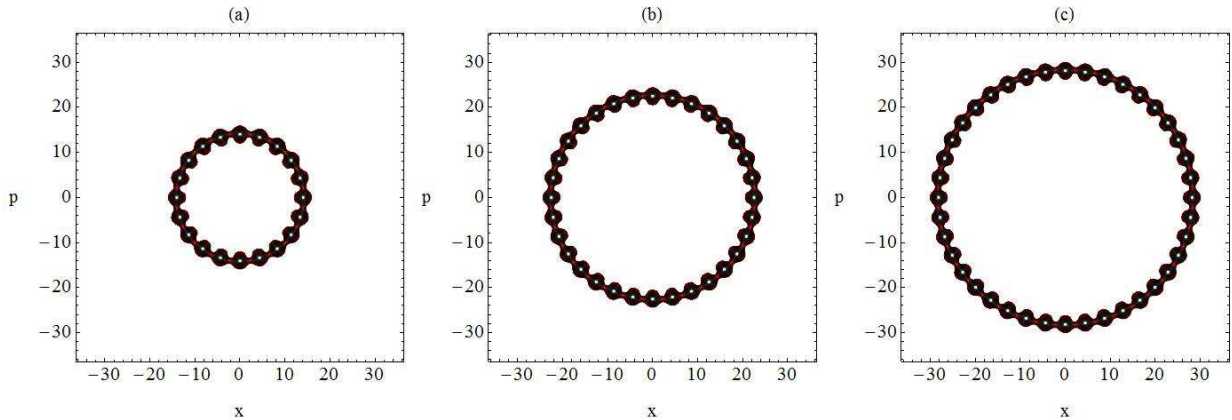}
\caption{Q-function in phase space for different $\alpha$: (a)
$\alpha=9.9$, $n=20$; (b) $\alpha=15.9$, $n=32$; and (c)
$\alpha=19.9$, $n=40$. We have plotted, maintaining the critical
ratio: $n/\alpha=\nu$, where the CSs are just touching each
other.}\label{QQ}
\end{figure*}

{\it Husimi-Q-Function}: So far, we have been discussing about the
OF, which is a result of superposition of $n$-CSs. However, it
becomes a natural question to ask: what is the physical situation
of the CSs on a circle for the critical ratio $n/\alpha=\nu$? We
have tried to explore the answer by calculating the
Husimi-Q-function. Q-function is a phase space distribution, which
is always positive and does not include any interference
structures. Hence, plotting the Q-function for a higher order
superposition is less time consuming. The Q-function is calculated
and then plotted in Fig.~\ref{QQ} for (a) $\alpha=9.9$, $n=20$;
(b) $\alpha=15.9$, $n=32$; and (c) $\alpha=19.9$, $n=40$. These
cases are evaluated for $n/\alpha=\nu$ and shows a striking
support to our intuition that the CSs are actually just touching
each other and starts becoming indistinguishable; thereby creating
a extended source of ring-shaped light of radius $\sqrt 2 \alpha$.
This established the required physical situation in the classical
optics to obtain ring shaped light source with constant angular
source strength.

Thus, the overlap function [Eq.~(\ref{bessel})] is the result of
superposition of $n$ CSs situated in a ring of radius $\sqrt 2
\alpha$. Hence. the superposition is analogous to the diffraction
pattern generated when light passes through the thin ring shaped
opening. The van Cittert-Zernike theorem \cite{Wolf} states that
diffraction problem is identical to the coherence problem and the
two problems result in the same mathematical formalism through the
quantity called complex degree of coherence. In this work, $n$
coherent states, symmetrically situated in circle produce
interference ripples in the center. This phase space interference
structures is known to manifest in the overlap function between
the original and displaced cat-states. We have shown that this
interference term is analogous to the diffraction pattern
resulting from the equivalent optical sources after proper scaling
of the parameters. Thus, the fact that the overlap between the cat
states and their shifted forms, is of the same form as the
diffraction pattern centered at one of the states, bears strong
resemblance to the van Cittert-Zernike theorem. Here the
normalized mutual coherence function $\gamma_{12}(0)$, for a ring
shaped opening with constant angular source strength, can be
written explicitly as,
\begin{eqnarray}
 \gamma_{12}(0)&=&\frac{\langle  E_1(t) E_2(t)^{*}\rangle_T}{\sqrt{\langle E_1(t) E_1(t)^{*}\rangle_T \langle  E_2(t) E_2(t)^{*}\rangle_T}}\nonumber\\
&=& J_0\left(\frac{2 \pi r_{0} |\vec{r_1}-\vec{r_2}|}{\lambda
R}\right).
\end{eqnarray}
$\gamma_{12}(0)$ actually signifies the complex degree of spatial
coherence of the two points at the same instant in time, when
fields arriving at the observation screen being $E_1(t)$ and
$E_2(t)$ respectively. $r_{0}$ is the radius of ring, $R$ is the
distance of the screen from the opening and
$|\vec{r_1}-\vec{r_2}|$ is the path difference between the points.

The suffix $T$ in the expectation value signifies the time average
according to the ergodic hypothesis. The above equation should be
compared with the OF for large $n$ (Eq.~\ref{bessel}), for unit
distance from the screen to the opening ($R=1$) and for
$[a,a^{\dagger}] = \pi \lambda^{-1}$ :
\begin{equation}
\langle
cat_{n,\alpha}|cat_{n,\alpha}^{\delta}\rangle=J_0\left(\frac{2 \pi
|\alpha| |\delta|}{\lambda}\right).
\end{equation}

\section{Conclusions}

In conclusion, the sensitivity of cat-like states to quantum
parameter estimation is studied for large number of constituent
CSs. The phase space support is explored for accessible parameter
ranges in realistic situation. We provide a quantitative estimate
of the phase space support for a given superposition. In the large
$n$ limit, the state OF, determining the orthogonality of cat and
displaced cat states, approaches the Bessel function. According to
the van Cittert-Zernike theorem, the coherence problem is
mathematically identical with the diffraction problem by complex
degree of coherence. The fact that the OF is having the same form
as the diffraction pattern results on the same expression of
normalized mutual coherence function for large $n$. This is
similar to the mutual coherence function of a circular ring, which
yields Bessel function of order zero, matching with the theorem of
van Cittert-Zernike. In addition, this work opens up several scope
for future studies: i) Mesoscopic superposition is a purely
quantum phenomenon and overlap function plays a very crucial role
in quantum parameter estimation. The overlap function, without
asymptotic limit, can be investigated for different relative
phases ($\theta_{\alpha}$ and $\theta_{\delta}$) of the CS and
displacement parameters, which will provide information in various
direction in phase space; ii) Studying mesoscopic superposition in
realistic quantum systems, which are not modelled by harmonic
oscillator CSs are quite nontrivial in general. Hence, it has
become a frequent practice to study harmonic oscillator system and
use that knowledge to investigate other solvable quantum
mechanical potentials; iii) Quantum sensitivity has become a very
fascinating area of research, where it is known that increasing
the value of the CS-parameter will make the system more and more
sensitive. However, it was completely unknown that the OF would
saturate to the Bessel function. This novel fact can be further
utilized for investigating various physical situations of
interest.

\section{Acknowledgment}

The author, R. Sharma acknowledges the support by National
Initiative on Undergraduate Science (NIUS) undertaken by
HBCSE-TIFR, Mumbai, India. Discussions with Prof. Vijay. A. Singh
of HBCSE are gratefully acknowledged.


\begin{thebibliography}{99}
\bibitem{normalization} Distingushibility is one of the essential criteria for observing
the interference structures of nonlocal superposition of localized
wave-packets. We have considered superposition of distinguishable
CSs. In some other context, if one needs to consider
indistinguishable CSs, which themselves overlapped each other, the
normalization will be $ \sqrt{1/ \sum_j^n\sum_k^n
Exp\left[-|\alpha|^2 (1-Exp[-(2i\pi(j-k))/n])\right]}$. For
distingushible states $j=k$ and it becomes $1/\sqrt{n}$, where $n$
is the total number of CSs.
\end{thebibliography}
\end{document}